\documentclass[pra,aps,twocolumn,nopacs,superscriptaddress,nofootinbib,reprint]{revtex4}
\usepackage{graphicx}  
\usepackage{dcolumn}  
\usepackage{bm}           
\usepackage{amsmath}
\usepackage{epsfig}
\usepackage{indentfirst}
\usepackage{psfrag}
\usepackage{subfigure}
\usepackage{amssymb}
\usepackage{tabularx,multirow}
\usepackage{color}
\usepackage[colorlinks,linkcolor=blue,citecolor=blue,urlcolor=blue,hyperindex,driverfallback=dvipdfm]{hyperref}
\usepackage[T1]{fontenc}
\def\rb{{\bf r}}  \def\Rb{{\bf R}}

\def\Eb{{\bf E}}      
  
   \def\kk{{{\bf k}_\parallel}}

\begin{document}
\title{Manipulating coherence of near-field thermal radiation in time-modulated systems}
\author{Renwen~Yu}
\affiliation{Department of Electrical Engineering, Ginzton Laboratory, Stanford University, Stanford, CA, USA}
\author{Shanhui~Fan}
\email{shanhui@stanford.edu}
\affiliation{Department of Electrical Engineering, Ginzton Laboratory, Stanford University, Stanford, CA, USA}
\begin{abstract}
We show that the spatial coherence of thermal radiation can be manipulated in time-modulated photonic systems supporting surface polaritons. We develop a fluctuational electrodynamics formalism for such systems to calculate the cross-spectral density tensor of the emitted thermal electromagnetic fields in the near-field regime. Our calculations indicate that, due to time-modulation, spatial coherence can be transferred between different frequencies, and correlations between different frequency components become possible. All these effects are unique to time-modulated systems. We also show that the decay rate of optical emitters can be controlled in the proximity of such time-modulated structure. Our findings open a promising avenue
toward coherence control in thermal radiation, dynamical thermal imaging, manipulating energy transfer among thermal or optical emitters, efficient near-field radiative cooling, and engineering spontaneous emission rates of
molecules.
\end{abstract}
\maketitle


Controlling thermal emission is crucial for many applications, such as passive radiative cooling \cite{RAZ14}, solar thermophotovoltaic energy conversion \cite{LBN14}, and incandescent lighting \cite{LTS11,IBC16}. Recent developments in near-field nanophotonics have provided new possibilities for enhanced radiative cooling \cite{GOP12}, thermophotovoltaic energy conversion \cite{LCG06,MLZ21,BZF09}, and imaging \cite{KMP05}. Near-field radiative heat transfer is also of fundamental importance to explore fluctuation physics at nanoscale \cite{CG99,JMM05,PV1971,VP07,BB14,OLF10,KHZ12,RSJ09,KSF15,SGS15,paper198,ZGZ17,SLL15,NSC08,PZB19,SZF16,BEK17,SOT14,BBJ11,AK22} in various aspects, including vacuum friction \cite{paper157,paper199}, persistent heat current at equilibrium \cite{ZF16}, and ultrafast radiative heat transfer \cite{paper286}. 

Most studies on controlling thermal emission are carried out with passive systems. But recent works have pointed to new effects when active systems are considered \cite{CSS15,ZFT19,BLF20,KPJ15,KLR15}. For example, in Ref.\ \cite{BLF20}, temporal modulation of the material permittivity is used to initiate an energy transfer between two optical resonances and to achieve active cooling in the far-field regime. Since, in the broader photonics context, temporal modulation has created promising new opportunities for photon management, such as frequency conversion \cite{SWF19,RSA19}, optical isolation \cite{YF09,SA17}, and optical temporal aiming \cite{PE20}, it should be of interest to further explore the consequence of temporal modulation on thermal radiation. 

A key characteristics of thermal radiation is its coherence property. Conventional thermal emitters can be well approximated by a blackbody, and its thermally emitted electromagnetic fields are incoherent in the far-field regime. In contrast, in the near-field regime, an enhanced coherence of thermal radiation from planar structures made of either noble metals or polar materials has been theoretically proposed \cite{CG99} and experimentally demonstrated \cite{DFC06}. This near-field coherence enhancement is due to the surface plasmon or phonon polaritons supported by these structures, and occurred only at the frequencies near the surface polariton resonances. Further nanophotonic structuring can transfer the enhanced coherence into the far field, leading to a coherent emission of light from thermal sources \cite{GCJ02}. However, all existing studies on coherence properties of thermal radiation consider only passive systems.

In this work, we study the spatial coherence of thermal radiation in time-modulated active systems. As an example, we consider a planar structure consisting of a plasmon or phonon polaritonic substrate and a dielectric thin film. We study the coherence properties of its thermally emitted electromagnetic fields in the near-field regime when the dielectric thin film is undergoing a permittivity modulation. Our study reveals several unique characteristics in the coherence properties. We show that the strong coherence can be observed at frequencies away from the surface polariton resonances, due to the transfer of coherence as induced by temporal modulation. Also, the temporally modulated system exhibits cross-frequency correlation, i.e., coherence between thermal radiations at different frequencies. None of these properties exist in passive structures. To study this system, we have developed a fluctuational electrodynamics formalism for time-modulated systems. Our formalism further reveals that the local density of states in the near field can be strongly affected due to temporal modulation. Thus, temporal modulation enables a novel pathway to control the decay rate of optical emitters.

\begin{figure}
\noindent \begin{centering}
\includegraphics[width=0.5\textwidth]{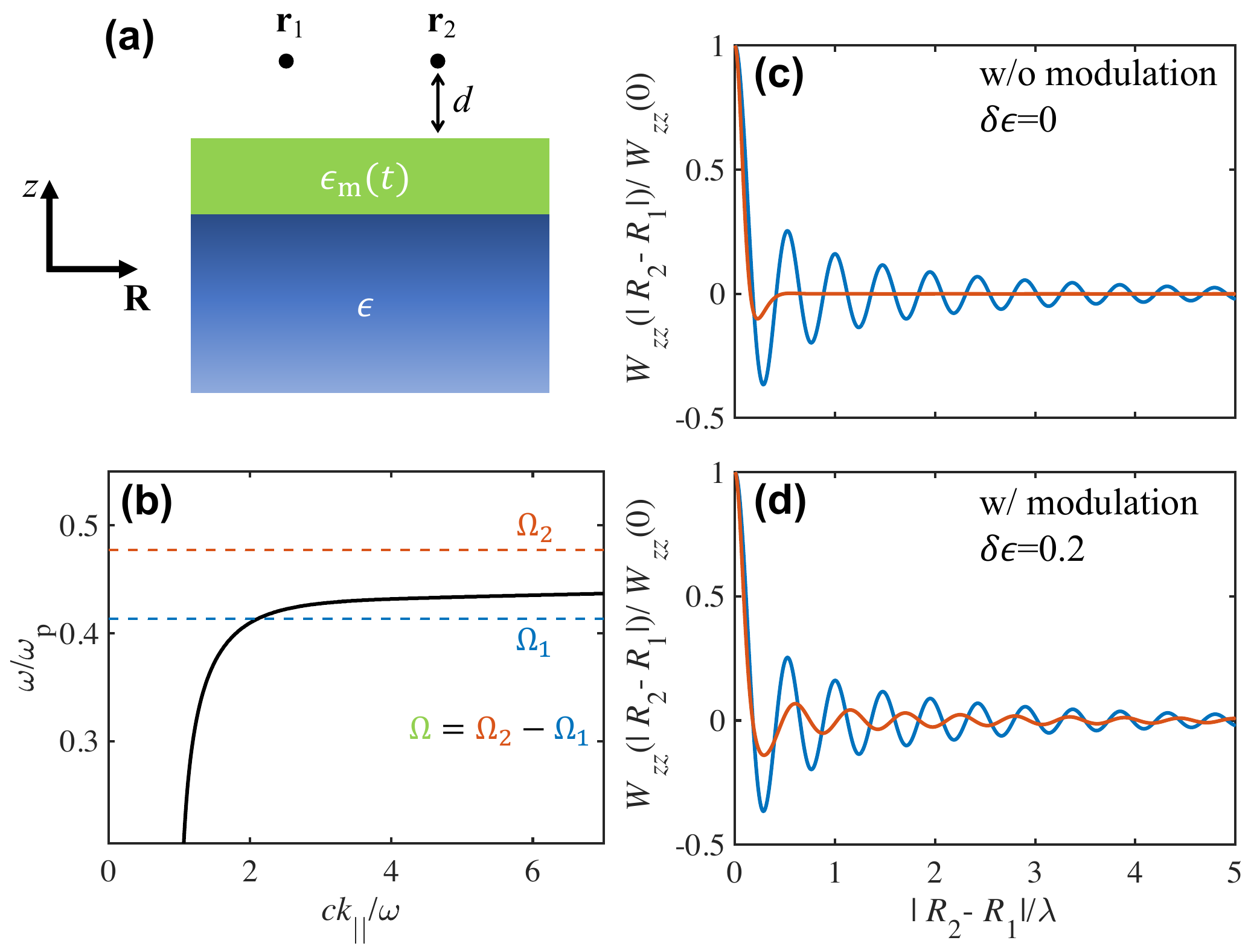}
\par\end{centering}
\caption{(a) Schematic of a photonic system composed of a time-modulated lossless layer (green region) on top of a substrate made of a Drude metal (blue region) with a plasma frequency $\omega_{\rm p}$. We consider the spatial correlation of the thermally emitted electric fields at two points $\rb_1$ and $\rb_2$ in vacuum. Both points are separated from the material surface by a vertical distance of $d=1\,\mu$m. (b) Dispersion relation (black solid curve) of the surface plasmon mode sustained in the system shown in panel (a). (c,d) Normalized element $W_{zz}$ of the cross-spectral density tensor as a function of the normalized lateral separation distance $|R_1-R_2|/\lambda$ between those two points shown in panel (a), without (c) or with (d, modulation strength $\delta\epsilon=0.2$) time modulation. Blue and red curves represent $W_{zz}$ evaluated at two frequencies $\Omega_1$ and $\Omega_2$ as labeled in panel (b). $\lambda=2\pi c/\omega$ is the light wavelength for the respective frequency in vacuum, and thus is different for the blue and red curves in panels (c,d).}
\label{Fig1}
\end{figure}

As the model system in this paper, in Fig.\ \ref{Fig1}(a), we consider a time-modulated lossless layer (green region) on top of a semi-infinite substrate composed of a Drude metal (blue region). The permittivity of the Drude metal is assumed to be $\epsilon(\omega)=1-\frac{\omega_p^2}{\omega(\omega+i\gamma)}$, with $\hbar \omega_p=0.13\,$eV and $\hbar \gamma=0.5\,$meV, and that of the time-modulated layer is $\epsilon_m(t)=\epsilon_s+\delta\epsilon\,{\rm cos}(\Omega t)$, with $\epsilon_s=4$ the static permittivity, $\delta\epsilon$ the modulation strength, and $\Omega$ the modulation frequency. Here, the modulated layer is assumed be lossless. Thus, all thermal radiation is sourced from the Drude metal. We assume that the top of the modulated layer is at $z=0$, and the thickness of the time-modulated layer is $1\,\mu$m. In the frequency domain, the thermally emitted electric fields ${\bf E}(\rb,\omega)$ in the vacuum region ($z>0$) can be obtained from the fluctuating currents ${\bf J}(\rb,\omega)$ residing in the Drude metal as
\begin{align}
	\Eb(\rb,\omega)=\frac{1}{2\pi}\int d \omega'\int d\rb' G(\rb,\rb',\omega,\omega')\cdot {\bf J}(\rb',\omega'), \label{eq1_old}
\end{align}
where $G(\rb,\rb',\omega,\omega')$ is the Green's function of the time-modulated system. In a static system, we have $G(\rb,\rb',\omega,\omega')=G(\rb,\rb',\omega)2\pi\delta(\omega-\omega')$. In contrast, for the time-modulated system with a single modulation frequency $\Omega$ as considered here, the Green's function is given by
$G(\rb,\rb', \omega, \omega') = \sum_n G(\rb, \rb', \omega, \omega') 2\pi\delta (\omega' - \omega_n)$ with $\omega_n=\omega+n\Omega$, and Eq.\ \ref{eq1_old} becomes
\begin{align}
	\Eb(\rb,\omega)=\sum_n\int d\rb' G(\rb,\rb',\omega,\omega_n)\cdot {\bf J}(\rb',\omega_n). \label{eq1}
\end{align}

 In this work, the focus is on the spatial coherence of thermally emitted fields. The physical quantity under investigation is the electric-field cross-spectral density tensor $W_{ij}$, given by \cite{MW95,CG99}
\begin{align}
	W_{ij}(\rb_1,\rb_2,\omega,\omega')=\left\langle E_{i}(\rb_1,\omega)E_{j}^{*}(\rb_2,\omega')\right\rangle, \label{eq2}
\end{align}
which characterizes the spatial correlation of the thermally emitted electric fields at two points $\rb_1$ and $\rb_2$ and at two frequencies $\omega$ and $\omega'$, respectively. The superscript * in Eq.\ \ref{eq2} denotes the complex conjugate operation. By using the fluctuation-dissipation theorem, from Eqs.\ \ref{eq1} and \ref{eq2}, the cross-spectral density tensor of the electric fields in the upper half space can be calculated as
\begin{align}
&	W_{ij}(\rb_1,\rb_2,\omega,\omega')= \sum_{n,m,v}H(\omega_n)\delta(\omega_n-\omega_m') \nonumber \\ 
&	\times \int d\rb'G_{iv}(\rb_1,\rb',\omega,\omega_n)G_{jv}^{*}(\rb_2,\rb',\omega',\omega_m'), \label{main1}
\end{align}
where $H(\omega_n)=4\pi\epsilon_0\omega_n{\rm Im}\left \{ \epsilon(\omega_n) \right\}\Theta(\omega_n)$ with $\Theta(\omega)=\hbar\omega\{\left[{\rm exp}(\hbar \omega/k_BT)-1 \right]^{-1}+1/2\}$. The integration over $\rb'$ is carried out inside the substrate (Drude metal here). We assume $T=300\,$K throughout this study. As seen in Eq.\ \ref{eq1}, an important aspect is that the thermal electromagnetic field at a frequency $\omega$ has contributions from sources at all frequencies $\omega_n$.
Consequently, as shown in Eq.\ \ref{main1}, electric fields at two frequencies $\omega$ and $\omega'$ can correlate with each other if $\omega-\omega' = l \Omega$, with $l$ being an integer. Such cross-frequency correlation for $l \neq 0$ arises from the photon frequency conversion as induced by the modulation with a frequency $\Omega$.  When Eq.\ \ref{main1} is applied to static systems, only the $n=m=0$ term contributes and thus $W_{ij}(\rb_1, \rb_2, \omega, \omega') \propto \delta(\omega-\omega') $. 

Because the system displayed in Fig.\ \ref{Fig1}(a) is translationally invariant in $x-y$ plane, Eq.\ \ref{main1} can be written in ${\bf k}_{\parallel}$-space as 
\begin{align}
&		W_{ij}(\rb_1,\rb_2,\omega,\omega')=\sum_{n,m,v}H(\omega_n)  \int dz' \int \frac{d\kk}{(2\pi)^2} \nonumber \\ 
&
\times	\overline G_{iv}(\kk,z_1,z',\omega,\omega_n) \overline G_{jv}^{*}(\kk,z_2,z',\omega',\omega_m') \nonumber \\
&	\times e^{i\kk\cdot(\Rb_1-\Rb_2)}\delta(\omega_n-\omega_m'), \label{main2}
\end{align}
where in-plane wave vectors $\kk=(k_x,k_y)$, in-plane coordinates $\Rb=(x,y)$, and $\overline G_{iv}$ represents the Fourier transform of $G_{iv}$ in $\kk$-space.
We have developed a fluctuational electrodynamics formalism for time-modulated systems to calculate $\overline G_{iv}$ and $W_{ij}$.

In Fig.\ \ref{Fig1}(b), we present the dispersion of the surface plasmon mode, supported in the system. We consider two frequencies $\Omega_1=2\pi\cdot13\,$THz and $\Omega_2=2\pi\cdot15\,$THz separated by $\Omega=\Omega_2-\Omega_1=2\pi\cdot2\,$THz. We use this value of $\Omega$ throughout this work unless otherwise specified. The structure supports a surface plasmon mode at $\Omega_1$ but not at $\Omega_2$, as indicated in Fig.\ \ref{Fig1}(b). Using Eq.\ \ref{main2}, we obtain $W_{zz}$ for $z_1=z_2=d$ in a plane at a subwavelength distance $d=1\,\mu$m (approximately 1/20 of the wavelength) above the modulated layer, as shown in Fig.\ \ref{Fig1}(a).

\begin{figure*}
\noindent \begin{centering}
\includegraphics[width=0.85\textwidth]{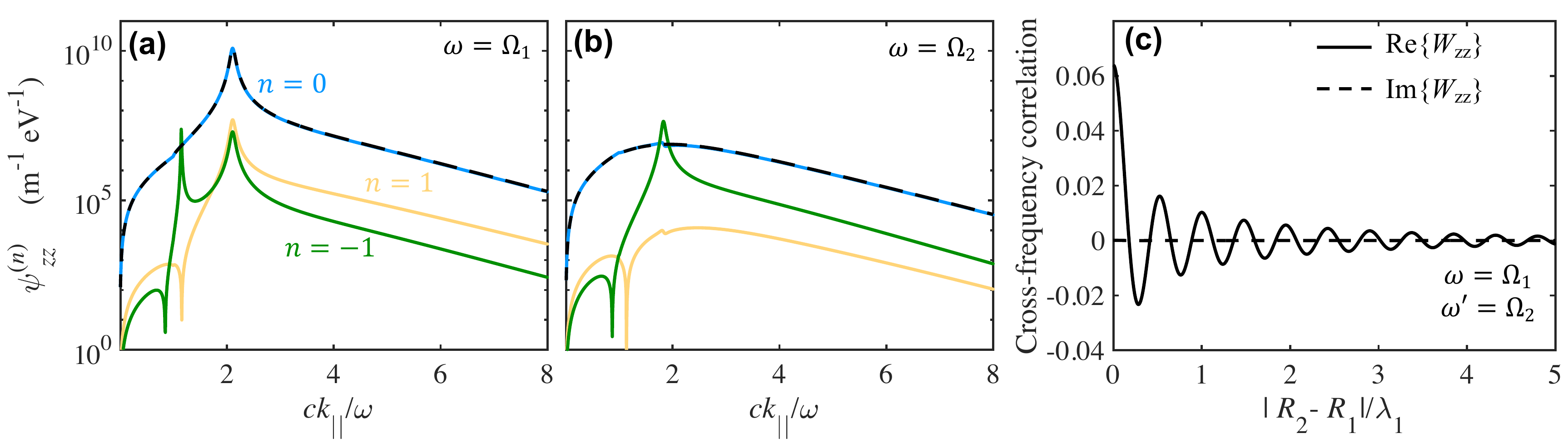}
\par\end{centering}
\caption{(a,b) Transition coefficient $\psi_{zz}^{(n)}$, defined in Eq.\ \ref{psi}, as a function of $k_{\parallel}$ at two fixed emitted field frequencies $\omega=\Omega_1$ (a) and $\omega=\Omega_2$ (b). Colored curves are for the system under the same time modulation as used in Fig.\ \ref{Fig1}(d). Black dashed curves are for the unmodulated system as used in Fig.\ \ref{Fig1}(c). (c) Real and imaginary parts of $W_{zz}(\omega=\Omega_1,\omega'=\Omega_2)$ as a function of normalized lateral separation distance under the same conditions as in Fig.\ \ref{Fig1}(d). Note that here $\lambda_1=2\pi c/\Omega_1$, and $W_{zz}$ is normalized to its value at $\Omega_1$ and $|R_1-R_2|=0$ without time modulation.}
\label{Fig2}
\end{figure*}

We first explore $W_{zz}$ for $\omega=\omega'$. In Fig.\ \ref{Fig1}(c), normalized $W_{zz}$ is plotted against the normalized in-plane separation distance $|R_1-R_2|/\lambda$ for the frequencies $\Omega_1$ and $\Omega_2$ without time modulation (i.e., $\delta\epsilon=0$). Because the surface plasmon dominates the electromagnetic response in the near-field regime, the correlation length is much larger at $\Omega_1$ where the surface plasmon mode is present (blue curve) than at $\Omega_2$ where no surface plasmon mode is available (red curve). The effect of such enhanced correlation length due to the presence of surface plasmons has been previously reported in Ref.\ \cite{CG99}. In contrast, when the time modulation is present [Fig.\ \ref{Fig1}(d)], we find the correlation length at $\Omega_2$ (red curve) is drastically enhanced as compared with the unmodulated case shown in Fig.\ \ref{Fig1}(c). In the meanwhile, a slightly decreased correlation length at $\Omega_1$ (blue curve) is observed [Fig.\ \ref{Fig1}(d)]. The results here indicate that the coherence is transferred from $\Omega_1$ to $\Omega_2$ due to time modulation.

In order to illustrate better the underlining mechanism behind results shown in Fig.\ \ref{Fig1}, based on Eq.\ \ref{main2}, we define 
\begin{align}
	&\psi_{zz}^{(n)}(\kk,d,\omega,\omega_n)=\frac{4\epsilon_0^2\omega_n^2}{\hbar\omega} 
	\sum_{v}\int dz'  \nonumber \\
	&\times \left|\overline G_{zv}(\kk,d,z',\omega,\omega_n) \right|^2. \label{psi}
\end{align}
$W_{zz}$ is related to $\psi_{zz}^{(n)}$ as 
\begin{align*}
& W_{zz}(\Rb_1,\Rb_2,\omega,\omega')=\frac{\hbar}{4\pi\epsilon_0} \sum_n{\rm Im}\left \{ \epsilon(\omega_n) \right\}\frac{\omega}{\omega_n}\Theta(\omega_n) \nonumber \\
& \times \int d\kk \psi_{zz}^{(n)}(\kk,d,\omega,\omega_n)e^{i\kk\cdot(\Rb_1-\Rb_2)} \delta(\omega-\omega')
\end{align*}	
for $\omega=\omega'$. Here, $\psi_{zz}^{(n)}$ can be associated with a spectrally resolved transition coefficient (in the unit of $\rm {m^{-1}\,eV^{-1}}$) from $\omega_n$ to $\omega$ for each $\kk$. 

In Fig.\ \ref{Fig2}(a), we plot $\psi_{zz}^{(n)}$ as a function of $k_{\parallel}$ for $n=-1$, 0, and 1 at a fixed emitted field frequency $\omega=\Omega_1$ under the same type of time modulation as used in Fig.\ \ref{Fig1}(d). The contributions to $W_{zz}$ from higher order ($|n|>1$) components are negligible. The dominant contribution is from the $n=0$ component (i.e., $\psi_{zz}^{(0)}$).  The spectrum of the $n=0$ component for the modulated system [blue curve in Fig.\ \ref{Fig2}(a)] is very similar to that of the unmodulated system [black dashed curve in Fig.\ \ref{Fig2}(a)]. Both spectra peaks at $k_{\parallel} = 2.1 \Omega_1/c$, which corresponds to the wave vector of the surface plasmon mode at $\Omega_1$, as shown in Fig.\ \ref{Fig1}(b). Thus both spectra are dominated by the contribution from the surface plasmon mode. As a result, in real space, the correlation function $W_{zz}$ oscillates at a period of approximately $2\pi/k_{\parallel}=\lambda/2.1$ for both the modulated and unmodulated cases, shown as the blue curve in Fig.\ \ref{Fig1}(c) and (d), respectively. The $n=-1$ component represents a transition from $\Omega_1-\Omega$ (thermal source frequency in the metal) to $\Omega_1$ (emitted field frequency in vacuum). From Fig.\ \ref{Fig1}(b), we observe that the system also supports a surface plasmon mode at $\Omega_1-\Omega$ with $k_{\parallel}\approx1.1\Omega_1/c$. Thus the spectrum of $n = -1$ component features two peaks around $k_\parallel = 1.1\Omega_1/c$ and $2.1 \Omega_1/c$ due to contributions from surface plasmon modes at $\Omega_1-\Omega$ and $\Omega_1$, respectively. Note that this transition between $\Omega_1-\Omega$ to $\Omega_1$ is not phase-matched. The modulation is spatially uniform parallel to the surface, and yet the wave vectors of the surface plasmon modes at $\Omega_1$ and $\Omega_1 - \Omega$ are different. Consequently, the strength of the transition is weak and thus the peak value of $\psi_{zz}^{(-1)}$ is much smaller than that of $\psi_{zz}^{(0)}$. The spectrum of $n=1$ component features only one peak around $k_\parallel = 2.1\Omega_1/c$ instead of two peaks because no surface plasmon mode is available at $\Omega_1+\Omega$ (i.e., $\Omega_2$). 

Significant differences in the behaviors of $\psi_{zz}^{(n)}$ can be found at $\omega=\Omega_2$. For the modulated system, the spectrum of its $n = 0$ component [blue curve in Fig.\ \ref{Fig2}(b)] is very similar to that of the unmodulated system [black dashed line in Fig.\ \ref{Fig2}(b)]. Also, both the $n = 0$ and $n = 1$ components do not exhibit any peaks since there is no surface plasmon mode at either $\Omega_2$ or $\Omega_2 + \Omega$. On the other hand, the $n = -1$ components exhibit one peak around $k_\parallel = 2.1\Omega_1/c=1.8\Omega_2/c$ because of the presence of the surface plasmon mode at $\Omega_2-\Omega = \Omega_1$. Also, because of the presence of the surface plasmon modes, the $n = -1$ component dominates over the $n=0$ and $n = 1$ components. Consequently, at $\omega = \Omega_2$, the modulated and the  unmodulated systems exhibit completely different behavior in their coherence properties. In the unmodulated system, there is no long-range correlation behavior [Fig.\ \ref{Fig1}(c), red curve] since there is no surface plasmon mode at $\Omega_2$. In the modulated system, there is a distinct long-range correlation behavior [Fig.\ \ref{Fig1}(d), red curve]. As a result, the correlation function exhibits a corresponding oscillation period around $\lambda/1.8$ in real space [Fig.\ \ref{Fig1}(d), red curve]. The results here indicate that time modulation can transfer coherence from one frequency to the others and hence can be used to drastically enhance the coherence of thermal radiation.

In Fig.\ \ref{Fig2}(c) we plot the cross-frequency correlation $W_{zz}(\Omega_1, \Omega_2)$, which describes the correlation of the electric fields at two different frequencies $\Omega_1$ and $\Omega_2$. The existence of such correlation for fields at two different frequencies $\Omega_1 \neq \Omega_2$ arises from the photonic transition as induced by dynamic modulation, and is a unique effect in time-modulated systems. Such cross-frequency correlation identically vanishes for passive systems. In Fig.\ \ref{Fig2}(c), a rather long correlation length can be seen, as a consequence of the coherence enhancement as provided by the surface plasmon mode at $\Omega_1$.  Note that $W_{zz}$ is now a complex quantity for $\omega \neq \omega'$, and in Fig.\ \ref{Fig2}(c) both its real and imaginary parts are plotted.

\begin{figure}
\noindent \begin{centering}
\includegraphics[width=0.5\textwidth]{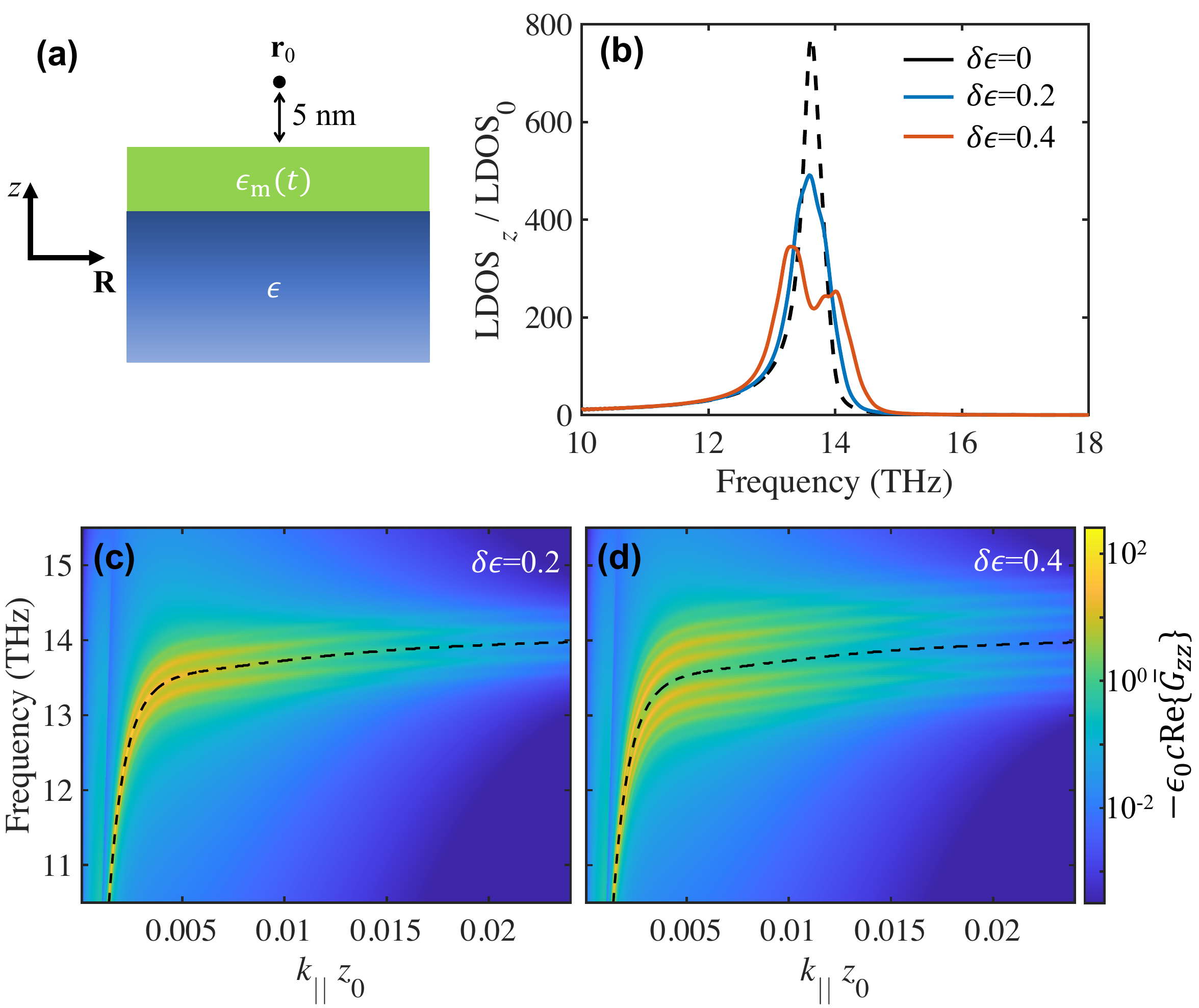}
\par\end{centering}
\caption{(a) The photonic system used here is the same as that shown in Fig.\ \ref{Fig1}(a) but with a modulation frequency of $\Omega=2\pi \cdot 0.2\,$THz. The $z$-polarized point dipole is placed at $\rb_0$ in vacuum ($z_0=5\,$nm away from the material surface). (b) LDOS spectra for different modulation strengths $\delta\epsilon$ (see legends inside). (c,d) We plot $-\epsilon_0c {\rm Re}\{ \overline G_{zz}(k_{\parallel},z_0,z_0,\omega,\omega)\}$ from Eq.\ \ref{LDOS} as a function of frequency and in-plane wave number $k_{\parallel}$ for $\delta\epsilon=0.2$ (c) and $\delta\epsilon=0.4$ (d). Black dashed curves represent the surface plasmon dispersion without time modulation.}
\label{Fig4}
\end{figure}

In addition to control the coherence properties, the time modulation can also be used to manipulate the local density of optical states (LDOS). The LDOS is closely related to the energy density of the emitted thermal electromagnetic fields. Moreover, by controlling the LDOS, the decay rate of an emitter in the vicinity of structure can be manipulated \cite{NH06,LKF14,paper352}. As shown in Fig.\ \ref{Fig4}(a), the system under investigation is the same as that used in Fig.\ \ref{Fig1}(a). Here we consider the contribution to the LDOS for the $z$-component of the electric field, at a location $\rb_0$ which is at $z_0 = 5\,$nm above the material-vacuum interface. Such a contribution is relevant for the spontaneous emission rate of a $z$-polarized dipole located at $\rb_0$. In time-modulated systems considered here, the LDOS can be calculated as
\begin{align}
\frac{{\rm LDOS}_z}{{\rm LDOS}_0}=-3{\rm Re}\left \{ \int \frac{k_{\parallel}dk_{\parallel}}{k_0^2} \epsilon_0c\,\overline G_{zz}(k_{\parallel},z_0,z_0,\omega,\omega)        \right\}. \label{LDOS}
\end{align}
We have normalized the LDOS to its value in free space ${\rm LDOS}_0=\omega^2/3\pi^2c^3$.

In Fig.\ \ref{Fig4}(b), LDOS spectra for different modulation strengths are plotted. Without time modulation, the spectrum consists of a single peak around 13.6\,THz (black dashed curve), as a result of the excitation of the surface plasmon resonance. In the presence of time modulation with $\delta\epsilon=0.2$, the spectrum consists of one broadened and damped peak (blue curve). With further increasing the modulation strength to $\delta\epsilon=0.4$, the spectrum further broadens and splits, and its peak amplitude further decreases (red curve). 

To analyze the LDOS spectra, we plot $-\epsilon_0c {\rm Re}\{ \overline G_{zz}(k_{\parallel},z_0,z_0,\omega,\omega)\}$, which appears in the integrand of Eq.\ \ref{LDOS}, as a function of frequency and in-plane wave number $k_{\parallel}$ in Fig.\ \ref{Fig4}(c) and (d). In the absence of modulation, this quantity should peak at the dispersion relation of the surface plasmon, as indicated by the black dashed line in Fig.\ \ref{Fig4}(c) and (d). In the presence of the modulation, this quantity peaks at the Floquet band structure of the system. For the case of $\delta \epsilon = 0.2$, in addition to the band at the dispersion relation of the surface plasmon, we see two prominent bands, with frequencies above and below the surface plasmon band respectively, due to the modulation. For the case of $\delta \epsilon = 0.4$, additional Floquet bands with frequencies above and below the surface plasmon band become visible. The broadening of the peaks of in the LDOS is directly related to the Floquet band structure of the system.

We remark that such time-modulation induced coherence transfer can be also achieved with a lower (sub-terahertz) modulation frequency and a smaller modulation strength (with $\delta\epsilon/\epsilon_{ s}$ around 0.01). The state-of-the-art electro-optical modulation with a response bandwidth approaching terahertz \cite{HNY12,MSY18,UHK19} can thus be exploited for this purpose. Moreover, in the present work we consider only direct photonic transition since our modulation is spatially uniform along the propagation direction of the surface plasmon. Additional opportunities may occur if one considers an indirect photonic transition \cite{YF09} where the modulation is of the form of $\delta \epsilon \,{\rm cos}(\Omega t - \beta x)$ with $\beta \neq 0$. Such a modulation may significantly enhance the coherence transition by coupling two surface plasmon modes at different frequencies together. Finally, the same concept and
mechanism can be used to achieve the coherence transfer in systems based on surface
phonon polaritons as well.


In summary, we have shown that the spatial coherence of thermal radiation can be manipulated by temporally modulating the permittivity of the dielectric layer in a polaritonic structure. This is initiated by the time-modulation induced transition of energy and coherence from multiple thermal source frequencies to the emitted thermal radiation frequency. As a result, correlations between two different radiation frequencies become possible in such system. Finally, we have also shown that the LDOS spectra can be controlled in the proximity of the time-modulated structure, due to the creation of the Floquet band structure in this system. Our findings open an exciting route
toward coherence control of radiation from nanoscale thermal sources, dynamical thermal imaging, manipulating energy transfer among remotely located thermal or optical emitters, enhanced near-field radiative cooling, and modulating spontaneous emission rates of
molecules.  

This work has been supported by a MURI program from the U. S. Army Research Office (Grant No. W911NF-19-1-0279).


\end{document}